# Extremely long-lived magnetic excitations in supported Fe chains

**J.P.Gauyacq[1] and N.Lorente[2,3]**

[1]Institut des Sciences Moléculaires d'Orsay (ISMO), CNRS, Univ. Paris-Sud, Université Paris-Saclay, F-91405 Orsay, France
[2] Centro de Fisica de Materiales CFM/MPC (CSIC-UPV/EHU),
Paseo Manuel de Lardizabal 5, 20018 Donostia-San Sebastian, Spain
[3] Donostia International Physics Center (DIPC),
Paseo Manuel de Lardizabal 4, 20018 Donostia-San Sebastian, Spain

## Abstract

We report on a theoretical study of the lifetime of the first excited state of spin chains made of an odd number of Fe atoms on $Cu_2N/Cu(100)$. Yan et al (Nat. Nanotech. **10**, 40 (2015)) recently observed very long lifetimes in the case of $Fe_3$ chains. We consider the decay of the first excited state induced by electron-hole pair creation in the substrate. For a finite magnetic field, the two lowest-lying states in the chain have a quasi-Néel state structure. Decay from one state to the other strongly depends on the degree of entanglement of the local spins in the chain. The entanglement in the chain accounts for the long lifetimes that increase exponentially with chain length. Despite their apparently very different properties, the behaviour of odd and even chains is governed by the same kind of phenomena, in particular entanglement effects. The present results account quite well for the lifetimes recently measured by Yan et al on $Fe_3$

**Introduction**

The recent development of low-T, high resolution, scanning tunnelling microscopy (STM) allows detailed investigations on spin systems at surfaces[1,2,3,4,5,6]. The study of adsorbed spin systems made of only a few atoms was a very significant breakthrough in the context of miniaturization of electronic devices. The possibility to characterize and modify spin systems at the atomic level opened the way towards detailed analysis of these systems, of their behaviour and in particular of their quantum aspects (see a review in [7]). The possibility to act on the spin variable of an adsorbate on a surface, i.e. to create local spin excitations, has been demonstrated; together with large magnetic anisotropies, this leads to the existence of well separated energy levels that could be used for logical devices. Indeed, atomic-scale logical devices have been assembled using atom manipulation techniques. A variety of magnetic nanostructures, and in particular of chains of magnetic atoms, have been built and analysed in STM experiments[8,9,10,11,12,13,14] that has triggered a series of associated or parallel theoretical studies[15,16,17,18,19,20,21].

Atomic spin systems adsorbed on surfaces a priori obey quantum dynamics, if decoherence effects are not too strong[22,23,24,25]; in addition, since adsorbed spin systems interact with the substrate, in particular with its thermal electron bath, magnetic excitations have finite lifetimes. For possible applications, it is of paramount importance to be able to determine the population lifetime of these spin excitations and their coherence time as well as to understand what the key parameters influencing these lifetimes are.

Few experiments have been devoted to the study of the lifetime of magnetic excitations in nano-structures at surfaces. We can mention the indirect measurement of population survival via STM-current saturation effects in the Mn/$Cu_2N$ system[12], and the direct lifetime measurement via STM experiments with high time resolution in even-numbered chains of Fe on $Cu_2N$/Cu(100)[13], and in $Fe_3$ on $Cu_2N$/Cu(100)[14]. Indeed, the different approaches were adapted to different time scales for the lifetime. In particular, the direct measurements, adapted to long time scales, revealed extremely long lifetimes, up to a few hours, for the Fe chains, that quickly increase in the even chain case as the chain length increases[14]. One can also mention measurements of excitation profiles[26,27,28]; indeed a finite lifetime leads to broadening of the peaks in an excitation spectrum. However, there are other sources of broadening besides experimental effects (e.g. temperature, decoherence ...), so that extracting a lifetime from a peak width can be a difficult procedure.

On the theoretical side, besides some studies involving adjustable parameters[29,30], only a few parameter-free studies have been reported including a perturbation calculation of electron scattering effects[31], as well as excitation-width calculations[32,33]; in both cases, a satisfying agreement with experimental data was reached.

Fe chains adsorbed on the $Cu_2N/Cu(100)$ surface present very appealing characteristics that make them choice systems for fundamental studies[12,13]. The nitride layer insulates the adsorbate from the metal substrate, thus limiting the flux of substrate electrons hitting the chain and reducing the decay rate of magnetic excitations. The chains can be described as a set of local spins interacting together (anti-ferromagnetic coupling, AFM) and with the substrate[12,13]. The magnetic anisotropy of each atomic spin is very large, so that the system is closer to an Ising chain than to a pure Heisenberg chain. As a consequence, the *classical equilibrium state of an even Fe chain* is a degenerate Néel state, with Fe atomic spins pointing parallel to the surface, in alternating directions along the chain. However, in a quantal view, the two Néel states are coupled (via exchange and/or transverse magnetic anisotropy), so that the ground state is a 50-50 mixture of the two Néel states. In the case of even chains, the detailed STM experimental study of Loth et al[13] showed that the two Néel states of the chain were observed and not the quantal ground state. It has been shown that inclusion of decoherence effects accounts for the observation of the classical Néel states[25,34]. Spontaneous flip between the two Néel states in the even Fe chains was also observed[13] and accounted for[25]; the flip rate quickly decreases with the chain length[13,25], so that long chains yield a good model system for nano-magnetic memories[13].

The case of *odd Fe chains* on $Cu_2N/Cu(100)$ is different, though it bears strong resemblances with the even Fe chain case. Indeed in both cases, the system is described by a set of local spins with a strong anisotropy and coupled by Heisenberg exchange. In a classical Ising model, the ground state is degenerate with two Néel-like states, with a non-vanishing spin projection and with Fe atomic spins pointing parallel to the surface, in alternating directions. When a magnetic field B is applied, the two Néel states of an odd chain split. The recent study by Yan et al[14] reported i) that a polarized magnetic STM tip very easily alters the system and ii) that for the $Fe_3$ chains, the upper Néel state has a very long lifetime, increasing roughly linearly with the applied magnetic field. It is the aim of the present work to compute the lifetime of the upper Néel state in a series of odd Fe chains on $Cu_2N$ using the method introduced by Novaes et al[31]. The decay of the magnetic excitation is induced by inelastic scattering of substrate electrons on the adsorbed chain, i.e. by electron-hole pair creation. This computation will lead to an analysis of the origin of the long lifetimes, as well as to a discussion of its dependence on the chain length.

## 2. Method

### 2.a Model description of the Fe chains

The odd Fe chains on Cu$_2$N/Cu(100) are described using the same model as for our earlier studies on spin chains[25,35,36]. It assumes that each Fe atom bears a local S=2 spin, with anisotropy terms, coupled by Heisenberg exchange. The effect of an applied B field is also included. The corresponding Hamiltonian reads:

$$H = \sum_{i=1}^{N-1} J \ \vec{S}_i . \vec{S}_{i+1} + \sum_i \left[ DS_{i,z}^2 + E(S_{i,x}^2 - S_{i,y}^2) \right] + \sum_i g\mu_B \vec{S}_i . \vec{B} \qquad (1)$$

where N is the total odd number of atoms in the chain (N=1-7 in the present study). $J$ is the Heisenberg coupling between neighbouring spins. $D$ and $E$ are the longitudinal and transverse magnetic anisotropies. The z-axis is along the easy magnetic axis, parallel to the surface. $\vec{B}$ is a macroscopic applied magnetic field; as in Yan et al experiment, it is aligned along the easy magnetic axis of the Fe atoms. $\mu_B$ is the Bohr magneton and g the gyromagnetic factor. The parameters in (1) for N>1 were taken from the spectroscopic study by Yan et al[14] on Fe$_3$ chains ($J$ = 1.15 meV, $D$ = -2.1(edge) and -3.6(center) meV and $E$ = 0.31 meV). For N=1, i.e. a single Fe atom on Cu$_2$N/Cu(100), we took the parameters ($D$ = -1.85 meV and $E$ = 0.31 meV) from the spectroscopic study of Hirjibehidin et al[2] on individual adsorbates. Below, the results of a test calculation are also presented, using the parameters determined by Bryant et al[17] ($D$ = -1.87 meV, $E$ = 0.31 meV and $J$ = 0.7 meV) from a spectroscopic study on Fe$_2$. Note that in all three cases[14,2,17], the parameter adjustment was on the spectroscopic part of the study, i.e. on the excitation energies in the system and not on the lifetimes. Starting from Hamiltonian (1), our computation of the lifetime is parameter free. The system described by (1) is not invariant by rotation; if $\vec{S}$ is the total spin of the system, $\vec{S}^2$ and $S_z$ are not good quantum numbers. Though, the total space splits in two uncoupled sub-spaces, corresponding to odd and even $S_z$ values. The eigenstates of the system are obtained by diagonalizing Hamiltonian (1) in a complete basis set formed by the tensorial product of local atomic spins (i.e. a single configuration of local spins). Note that the local spins are independent in a single configuration description; however, diagonalizing Hamiltonian (1) will generate correlated wave-functions for which the local spins are entangled.

Because of the large negative longitudinal anisotropy, $D$, *for a finite B field*, the two lowest eigenstates ($\mid N_1 \rangle$ and $\mid N_2 \rangle$) are almost two Néel states with Fe atomic spins pointing parallel to the surface, in alternating directions (see below the case of a vanishing B field). The projection of the total spin on the z-axis is almost equal to +2 and -2 for the two states. Because of the existence of $J$ and $E$ terms, the two pure Néel states are coupled together via a high order indirect coupling, so that the result of the diagonalization of (1) does not yield exactly Néel states. When B is increased, the two states split almost linearly and the present study, as well as the earlier experimental study by Yan et al[14], is devoted to the decay of $\mid N_2 \rangle$, the upper quasi-Néel state, to the lower, $\mid N_1 \rangle$.

## 2.b Computation of the decay rate of the excited Néel state

The decay of the excited state is induced by inelastic scattering of a substrate electron by an atom of the chain that leads to the change of the magnetic state of the whole chain (electron-hole pair creation). The collision between an electron and an atom can only change the atomic spin by one unit of angular momentum, $\mid \Delta M \mid = \pm 1$. In the present case, to flip between the two quasi-Néel states requires $\mid \Delta M \mid \approx \pm 4$ on each of the atoms in the chain. As a consequence, if the $\mid N_1 \rangle$ and $\mid N_2 \rangle$ states were pure Néel states, no transition between them could be induced by an electron collision. However, with Hamiltonian (1), the projection of the total angular momentum on the z-axis is not a quantum number and the $\mid N_1 \rangle$ and $\mid N_2 \rangle$ states are mixtures of many states. In other words, the spin chain is correlated, it is not represented by a single configuration of local spins and the various spins along the chain are entangled. This can be linked to the large entanglement of the local spins in the ground state of pure Heisenberg chains[37,38], which strongly influences excitation processes in spin chains by tunnelling electron[36]. In the odd chains of Fe case, an electron collision on a single atom in the chain can flip all the spins in the chain; however, the probability for a global flip of the chain is very weak, because of the weakness of the atomic spin entanglement, or in other words since the $\mid N_1 \rangle$ and $\mid N_2 \rangle$ states are almost Néel states due to the large $D$ term in Eq. (1).

The lifetime $\tau$ of the excited state $\mid N_2 \rangle$ is the inverse of $\Gamma$, the decay rate from $\mid N_2 \rangle$ to $\mid N_1 \rangle$. The decay corresponds to the super-elastic (final electron energy larger than its initial energy) scattering of substrate electrons by the adsorbate. The decay rate for a vanishing temperature can be expressed by the $T$ transition matrix for electron scattering ($j \rightarrow f$) by one of the atoms in the chain as (see a discussion of the decay rate by electron-hole pair creation in[39] and the application to spin decay in[7,31,40]):

$$\frac{1}{\tau} = \Gamma = \frac{2\pi}{\hbar} \frac{\delta\Omega}{\sum_{k_j, k_f, m_j, m_f}} \left| \langle k_f, m_f, N_1 | T | k_j, m_j, N_2 \rangle \right|^2 \delta(\varepsilon_j - \varepsilon_f) \delta(\varepsilon_j - E_F) \qquad (2)$$

where $\delta\Omega = E_2 - E_1$ is the energy difference between $|N_2\rangle$ and $|N_1\rangle$ states. The total energy is $E_T = E_2 + \varepsilon_j = E_1 + \varepsilon_f$. The initial state and final states of the substrate electrons are labelled by their energy $\varepsilon_j$ and $\varepsilon_f$, their wave number, $k_j$ and $k_f$, and by their spin projections on the quantization axis, $m_j$ and $m_f$. Equation (2) is derived under two assumptions (see details in ref.[31,39,40]): i) the system temperature is assumed to be very small and ii) the $T$ transition matrix elements are assumed to be constant in the small energy interval involved in the decay process. Inelastic scattering of the substrate electrons with the chain of magnetic atoms is assumed to only occur with one atom of the chain at a time and scattering is fast so that a sudden approximation, neglecting the magnetic Hamiltonian during the scattering process can be used. The $T$ transition matrix is then diagonal in the basis set formed by the eigen-states of $\vec{S}_T^2$ and $S_{T,z}$ (quantum numbers, $S_T$ and $M_T$), where $\vec{S}_T$ is the total spin of the scattering atom+electron system[7,41]. In the present case of Fe atoms, $S_T$ can be equal to 3/2 or 5/2. In our earlier study on Mn/Cu$_2$N/Cu(100), it appeared that both $S_T$ values contribute to scattering and we included both in a statistical way[31]. Similarly, below, we assume that the two $S_T$ values contribute to the decay independently, in a statistical way. Following references[7,31,40], one can show that the decay rate reduces to :

$$\frac{1}{\tau} = \Gamma = T_{total}(E_F) \frac{\delta\Omega}{h} P_{Spin}(N_2 \rightarrow N_1) \qquad (3)$$

where $T_{Total}(E_F)/h$ is the total electron flux hitting a chain atom per unit of energy and per unit of time. $T_{Total}(E_F)$ typically amounts[31] to 1 (recent calculations by Novaes, Lorente and Gauyacq on individual Fe adatoms on Cu$_2$N/Cu(100) yielded $T_{Total}(E_F)$ values in the same range as those obtained for Mn adsorbates in ref.[31]). Below, we take this flux factor equal to 1. $P_{Spin}(N_2 \rightarrow N_1)$ is the probability for a transition from $N_2$ to $N_1$ during the collision of an electron and an atom of the chain, averaged over the two $S_T$ values.

Expression (3) corresponds to the decay induced by electron collisions on a single atom of the chain. Below, we assume that the electron collisions on the different atoms contribute to the decay in an independent way and so we sum the contributions (3) from the different atoms (those are slightly different depending on the different positions of the atoms in the chain).

## 3. Lifetimes

### 3.a Comparison between the different chains

Figure 1 presents the lifetime of the first excited state (state $|N_2\rangle$) of Fe chains as a function of the applied macroscopic magnetic field $B$ for various chain lengths (N=1, 3, 5 and 7). The behaviour of $\tau(B)$ is similar for the various chains. In the moderate B field range (1-2 T) where the first excited state, $|N_2\rangle$, has a quasi-Néel state character, the lifetime appears to be roughly linear with B for all chains. In that range, the energy defect of the decay, $\delta\Omega$, varies linearly with B (Zeeman term). In contrast, the probability entering in equation (3) depends on the spin entanglement in the chain, i.e. on the mixing between the two Néel states. Indirect high order terms in $J$ and $E$ are independent of B and since the energy difference between the two Néel states varies like B, this yield a mixing probability that varies like $1/B^2$. Following equation (3) this leads to a variation of the lifetime linear in B. In the high B range in Fig.1, the Néel state $|N_2\rangle$ comes closer to high lying states and mixes with them; it thus further loses its quasi-Néel state character; these mixings facilitate the decay and the state becomes gradually more unstable. The sharp drop in the Fe case around 12.5 T is due to a crossing between the quasi-Néel state and another state of different spin symmetry.

As the main striking feature, the lifetime is found to increase very rapidly with the chain length, reaching very long times, up to the hour range in the $Fe_7$ case. The lifetime increases roughly by a factor 20000-30000, when adding two Fe to the chain. Decay corresponds to the flip of all the atomic spins when an electron collides with a single atom and thus requires entanglement of all the atomic spins. The drop of the decay rate with the chain length in Fig.1 is correlated to the drop of entanglement of the atomic spins. In other words, the $N_1$ and $N_2$ states have a stronger Néel state character as the length increases; indeed, mixing between the two pure Néel states requires higher order couplings involving $J$ and/or $E$ as the length increases. Because of this, one can expect an exponential drop of the decay rate with the chain length (see further details below in the low-B section).

The case of a single Fe adsorbate could appear as different from the other ones, although its lifetime fits well in the evolution observed on the chains. Indeed, entanglement of different atomic spins cannot be invoked in this case. As discussed in Hirjibehedin al[2], at

B=0, the two lowest states correspond to an equal mixing of the $S_z$=+2 and -2 of Fe with a small contribution from the $S_z$=0 state. This structure is driven by the magnetic anisotropies $D$ and $E$. When a B field is applied, the Zeeman term is able to uncouple the structure governed by the anisotropies and yields states that become more and more pure $S_z$ states as B increases. At finite B, the transition between pure $S_z$=+2 and -2 states is impossible via collision with an electron; however, since there is some correlation induced by the transverse anisotropy (the system state is a superposition of different $S_z$ states) a weak transition is possible, leading to the lifetime displayed in Fig.1 that increases as the applied field increases. The origin of long lifetimes in systems with large negative $D$ terms, like single Fe atom, has already been discussed in [42]. We conclude that the role played by the $J$ and $E$ coupling terms in the chain case reduces to that of the $E$ term alone in the Fe case; in all cases, decay proceeds via the correlations in the system: in the Fe case, the system state is not exactly a $S_z$ eigenstate and in the chain case, the system state is not exactly a single configuration of atomic spin states.

**3.b Low B-field limit**

The situation in the limit of vanishingly small B field is different from the one outlined above. It is illustrated in Figure 2, which shows the lifetime of the first excited state at a function of B on a double logarithmic scale to enhance the behaviour at extremely low B.

For B=0, the two Néel states are degenerated and coupled by indirect high order $J$ and $E$ terms (only the $E$ term in the single Fe atom case). The two lowest eigenstates of Hamiltonian (1) are then equal mixtures of the two Néel states with small admixtures of higher lying states. The energy difference between the two low-lying states is very small and equal to twice V, the effective coupling between the two Néel states. The situation of the odd chains at B=0 is thus very close to that of the even chains[25, 35].

When a very small, finite, B field is applied, the Zeeman term splits the two quasi Néel states when it is able to overcome the V coupling and the system is in the situation described in the previous section. The lifetime of the first excited state is then constant at very small B and then switches to the linear behaviour discussed in the previous section (see Figure 2). The transition between the two regimes is similar in the different chains, although it appears in very different B ranges. Indeed, the switch occurs in the region where the Zeeman term is of the order of V, the effective coupling between the two Néel states, i.e. when it is large enough to decouple the two Néel states. The effective coupling V is due to higher and higher order indirect couplings as the chain length is increased and it decreases roughly exponentially with the length. As a consequence, the low B behaviour occurs at lower B field as the chain length is increased, as seen in Fig.2. Actually, for long chains, the low field regime might be difficult if not impossible to reach; any small stray magnetic field or more generally any small perturbation is able to uncouple the B=0 structure and leads to the regime discussed in the section 3a.

The effective coupling, V, can be obtained as half the energy splitting of the two eigenstates at B=0. It is shown in figure 3 as a function of the chain length. The present results (red squares) are seen to decrease roughly exponentially with the chain length, as expected for an indirect process involving higher order terms as N increases. These results were obtained with the set of magnetic parameters determined from a spectroscopic study by Yan et al for $Fe_3$[14] chains. Results obtained for odd and even chains with another set of magnetic parameters (Bryant et al[17] set obtained from a spectroscopic study of $Fe_2$ and used in Ref. [25] for the study on even chains) are also shown. It appears that the V coupling for odd and even chains with Bryant et al parameters follows the same exponential behaviour, so that again, odd and even chains, although they seem to behave qualitatively differently at first sight, actually exhibit the same kind of couplings. It also appears that the different sets of parameters lead to different results: they are very close for the short chains, e.g. for $Fe_3$, but become significantly different for long chains; this is simply a consequence of the high order of the indirect coupling scheme that leads to the V coupling: the higher is the order, the more sensitive the V coupling is on small variations in the magnetic parameters. Similarly, for very long chains, the lifetime is sensitive to the precise choice of magnetic parameters.

One can stress that for a vanishing B and a long chain, one has two low lying eigenstates that interact with the underlying continuum of substrate electrons. Elastic collision of substrate electrons by the chain brings decoherence into the chain quantal system. Again, the situation is identical to that met in even chains (see a discussion in [25]). In most realistic situations, the above computed lifetime for B=0 in long chains is not a meaningful quantity. As for even chains[13], except for extremely low T, an STM observation on long chains would not observe the system quantal ground state but would find an equal statistical population for the two Néel states and if the system is initially in one Néel states, it will relax slowly toward a classical equilibrium, with equal incoherent populations of the two Néel states[25].

### 3.c Comparison with Yan et al data

Yan et al[14] recently reported measurements of the first excited state ($|N_2\rangle$) lifetime in $Fe_3$ chains and Figure 4 compares their results with the present ones. Very interestingly, Yan et al[14] found different values for the state lifetime when the STM tip was placed on the central or on an edge atom of the chain. They showed that this was due to the perturbation of the chain by the polarized tip that disappears when the tip is moved at larger distances from the chain. Only for a large tip-chain distance, can they measure the intrinsic lifetime. Figure 4 presents the two lifetimes measured by Yan et al[Error! Marcador no definido.] on the central and edge atom of $Fe_3$ as a function of the applied B field for a fixed tip-chain distance as well as the 'asymptotic' intrinsic value at B=2T found for a large tip-chain distance. The calculated results (blue curve) correspond to an unperturbed chain and should then be compared to some average of the two sets of experimental data and/or to the 'asymptotic' point at 2T. The present results appear to reproduce the B-variation observed experimentally; however, they roughly underestimate the experimental data by a factor 2.8, i.e. the discrepancy is of the same order as the one found in our earlier study[31] on single Mn adsorbates on $Cu_2N/Cu(100)$. Possible inaccuracies in the theoretical or experimental studies could account for this. However, one can also stress the sensitivity of the present results to the magnetic parameters in Hamiltonian (1). To illustrate this sensitivity, Figure 4 also presents the results obtained with a 6.15 % change of the $J$, $E$ and $D$ parameters ($E$ and $J$ are lowered and $D$ is increased, i.e. the changes bring the system closer to a pure Ising model with smaller entanglement in the chain and so a longer lifetime). The 6.15 % change is chosen to bring the computed results in agreement with the unperturbed experimental data at 2T (the three parameters, $J$, $D$ and $E$, act in a similar way and agreement with experiment can also be found when modifying only one of the three parameters). This small change is to be compared with the accuracy of the spectroscopic fit reported by Yan et al[14] on the energies which amounts to around 5% for $D$ and 10% for $J$. As a consequence, the discrepancy between theory and experiment seen in Fig.4 is tentatively attributed to inaccuracies in the parameters of the magnetic Hamiltonian (1).

Yan et al[14] also showed that the perturbation on the $Fe_3$ chain induced by the polarized STM tip could be represented by an effective magnetic field, $B_{Local}$, only active on the atom under the tip. The effective local field is linked to the tip-chain distance, z, by:

$$B_{Local} = B_0 \ (exp(-\gamma z)-1) \qquad (4)$$

where $B_0$ and $\gamma$ are parameters characterizing the tip-atom coupling strength. z is equal to zero at a distance where the tip perturbation is not measurable and decreases when the tip approaches the surface. We introduced this $B_{Local}$ term in Hamiltonian (1) and the corresponding results for $Fe_3$ (not shown here) reproduce those of Yan et al modelling[14] within a multiplicative factor (note that Yan et al modelling involves a general scaling of the lifetime). In Figure 5, we present the results for the $Fe_5$ chain. It shows the lifetime of the first excited state as a function of the local magnetic field acting on one of the three different atoms in the chain, i.e. the lifetime measured above one of the atoms in the chain when the tip-chain distance is varied. In the $Fe_5$ chain, three different sites exist: central, intermediate and edge. The local field, $B_{Local}$, is parallel to the applied global field, B (B = 2T in Fig.5), and pointing in the same direction. As a consequence the $B_{Local}$ Zeeman term acting on an intermediate atom decreases the energy difference between the two lowest lying states. As a function of $B_{Local}$, this leads to a sharply avoided crossing in the region close to where $B_{Local}$ = B. The situation is then similar to the threshold region discussed above and the lifetime undergoes a sharp minimum. The two regions above and below the critical region ($B_{Local} \approx B$) are analogous, except for an interchange between the ground state and the first excited state. For the other two sites, the energy difference between the two lowest states increases when a $B_{Local}$ is applied and the lifetime of the first excited state increases regularly. The situation for $Fe_5$ is then very similar to that of $Fe_3$ discussed by Yan et al[14]. However, we can stress that in Fig.5, the lifetime in sites 1 and 3 are quasi-equal, although these are non-symmetric positions (centre and edge sites). So, for $Fe_5$, the lifetime values measured on the atoms oscillate along the chain; in particular for a moderate $B_{Local}$ perturbation, the lifetime measured above similarly polarized atoms is basically the same, independently of the actual position of the atom along the chain.

## 4. Concluding summary

We have reported on a theoretical calculation of the lifetime of the first excited state in odd Fe chains adsorbed on $Cu_2N/Cu(100)$. The decay is induced by inelastic scattering of substrate electrons on the chain (electron-hole pair creation). The lifetime is found to be very long, increasing exponentially with the chain length, for $Fe_7$ the lifetime already reaches the hour range.

The two lowest lying states of an odd chain for a finite B have a quasi-Néel state structure. Switching from one spin state to the other under a single-electron interaction is then very difficult, depending on the correlated structure of the chain, i.e. on the degree of mixing between the two Néel configurations in the system eigenstates. The entanglement of the atomic spins in the chain is then responsible for the possibility of decay by electron-hole pair creation. As discussed earlier[42], a negative longitudinal anisotropy, $D$, favours long lifetimes in a single adsorbate. In the odd Fe chain case, this effect is much enhanced by the weakness of the entanglement of the local spins and accounts for the extremely long lifetimes and their fast increase with the chain length.

The present results appear to reproduce reasonably well the experimental data of Yan et al on Fe$_3$[14], allowing for a global factor of 2.8. The difference is attributed to an insufficient accuracy of the magnetic structure parameters of the chain. The very long lifetimes in the system are very sensitive to small variations in the magnetic structure parameters (exchange coupling and anisotropies). Computation of magnetic structure parameters have been performed via configuration interaction or DFT approaches for a variety of nanostructures at surfaces systems[15,18,19,20,43,44,45,46,47]. Further studies by ab initio approaches on the present system could be highly conclusive, despite the current difficulties in accurately computing magnetic structure parameters[48,49].

It turns out that odd and even Fe chains on Cu$_2$N/Cu(100) have very close magnetic structures, the basic interactions at play, in particular a weak entanglement of the local spins, being the same. In the absence of an applied magnetic field, both odd and even chains exhibit two closely spaced states, strongly influenced by decoherence. However, a small magnetic field is able to split the two states in the odd chain case, leading to the possibility of a very long lived state as observed by Yan et al[14] and computed here.

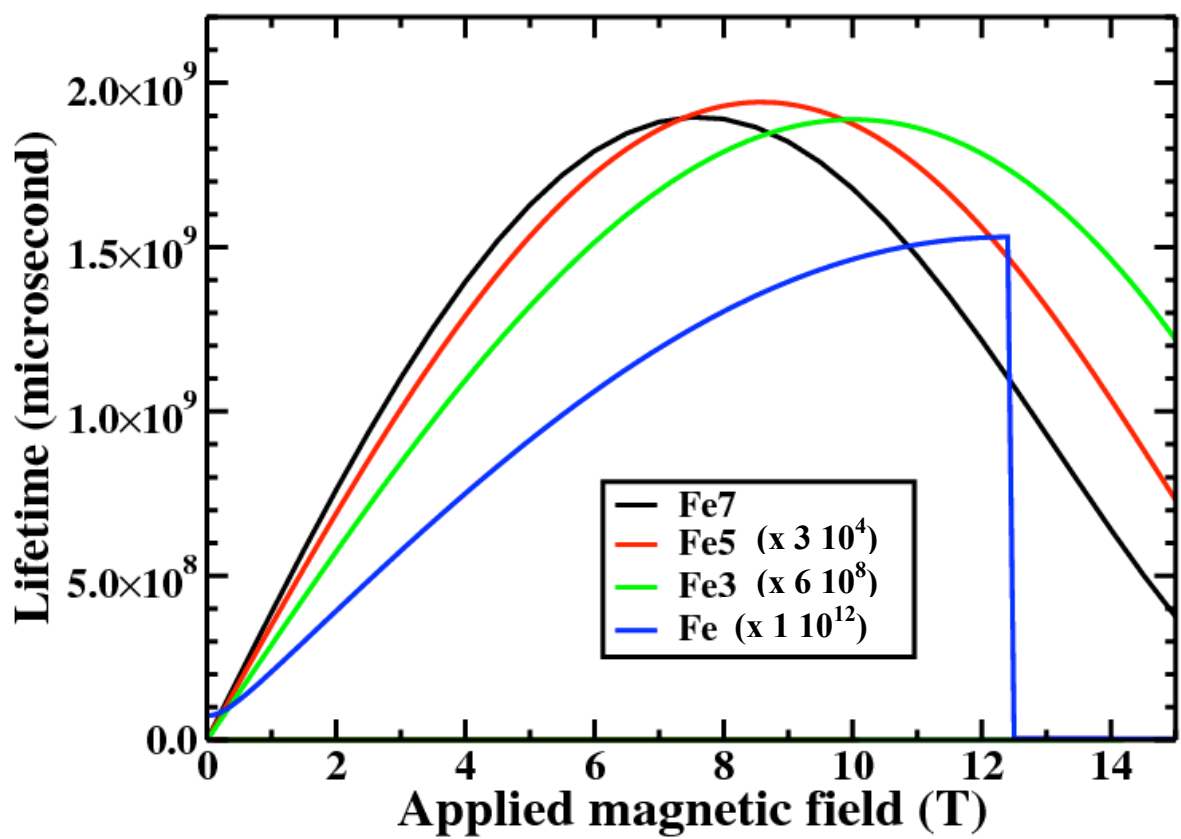

**Figure 1**: Lifetime in microsecond of the first excited state (quasi-Néel state $N_2$) of odd chains of Fe atoms adsorbed on a $Cu_2N/Cu(100)$ surface. The lifetime is plotted as a function of the macroscopic magnetic field (in T) applied to the system. Four different chain lengths are presented: N=1 (blue), N=3 (green), N=5 (red) and N=7 (black). Note the scaling factors on some of the curves.

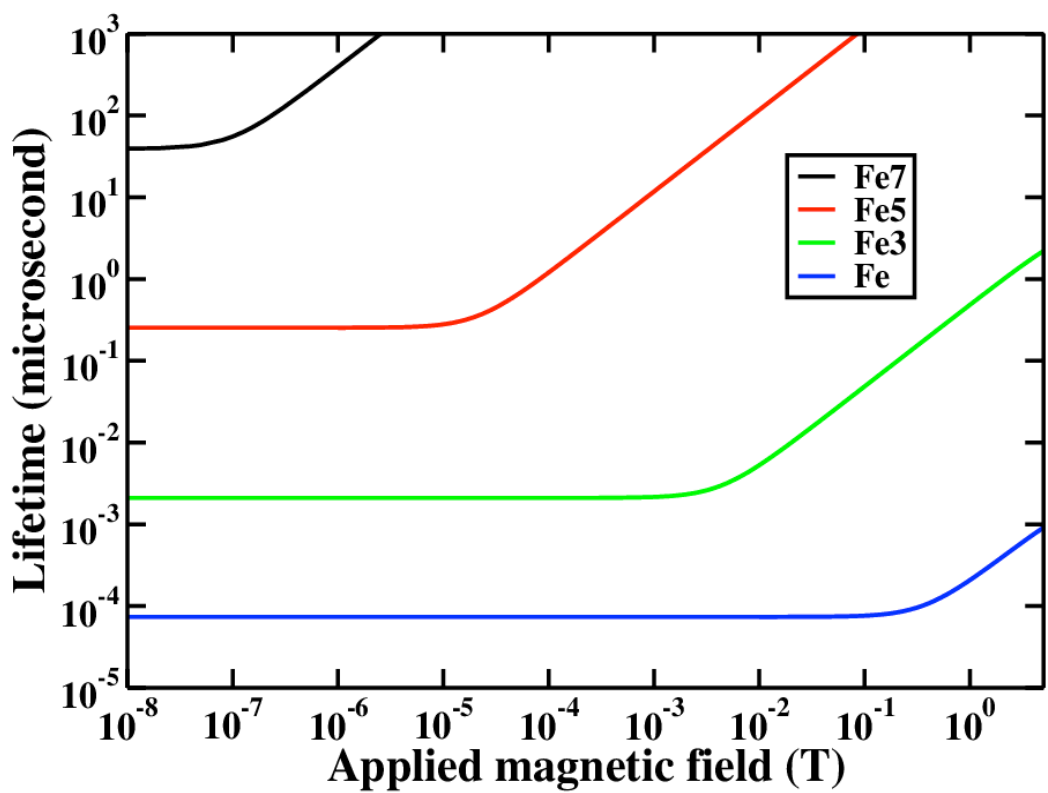

**Figure 2**: Lifetime in microsecond of the first excited state (quasi-Néel state $N_2$) of odd chains of Fe atoms adsorbed on a $Cu_2N/Cu(100)$ surface. The lifetime is plotted as a function of the macroscopic magnetic field (in T) applied to the system. Four different chain lengths are presented: N=1 (blue), N=3 (green), N=5 (red) and N=7 (black). A double logarithmic scale is used to stress the change of regime at very low B field.

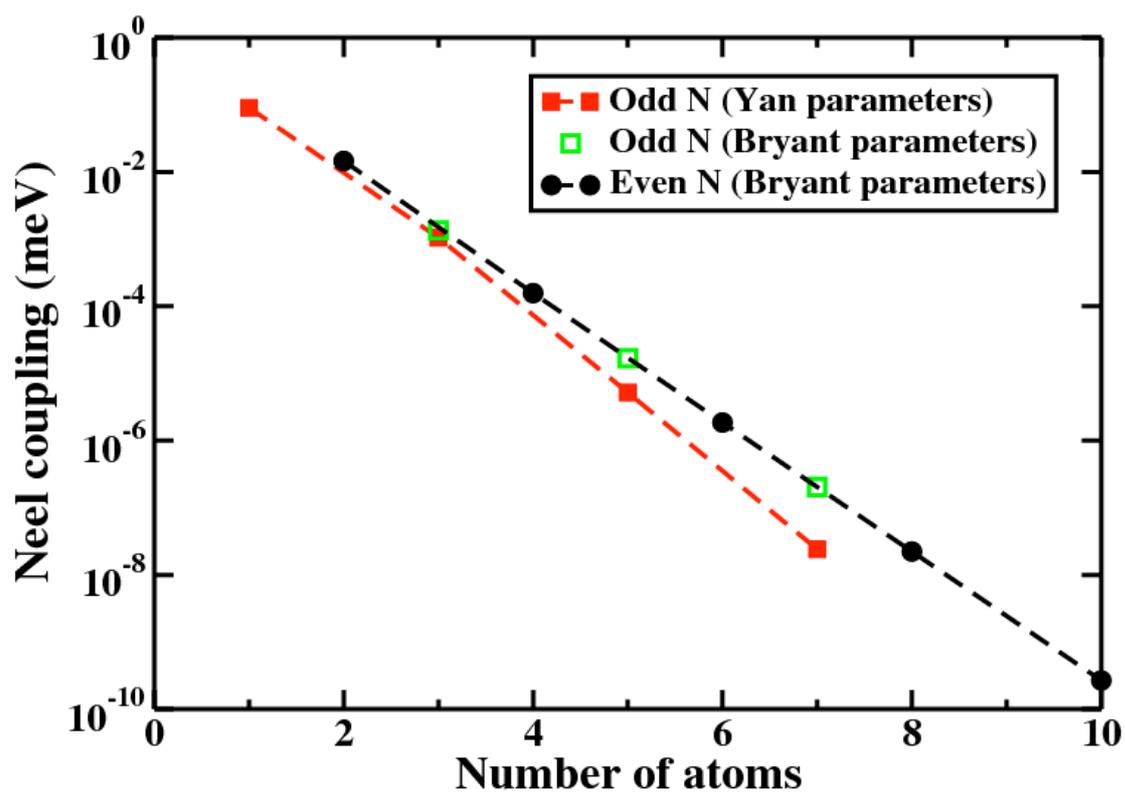

Figure 3: Effective coupling, V, between the two Néel states as a function of the chain length. Several sets of results are shown: present results for odd chains (red squares) and results of tests for odd and even chains obtained with Bryant et al[17] set of parameters (green squares and black dots).

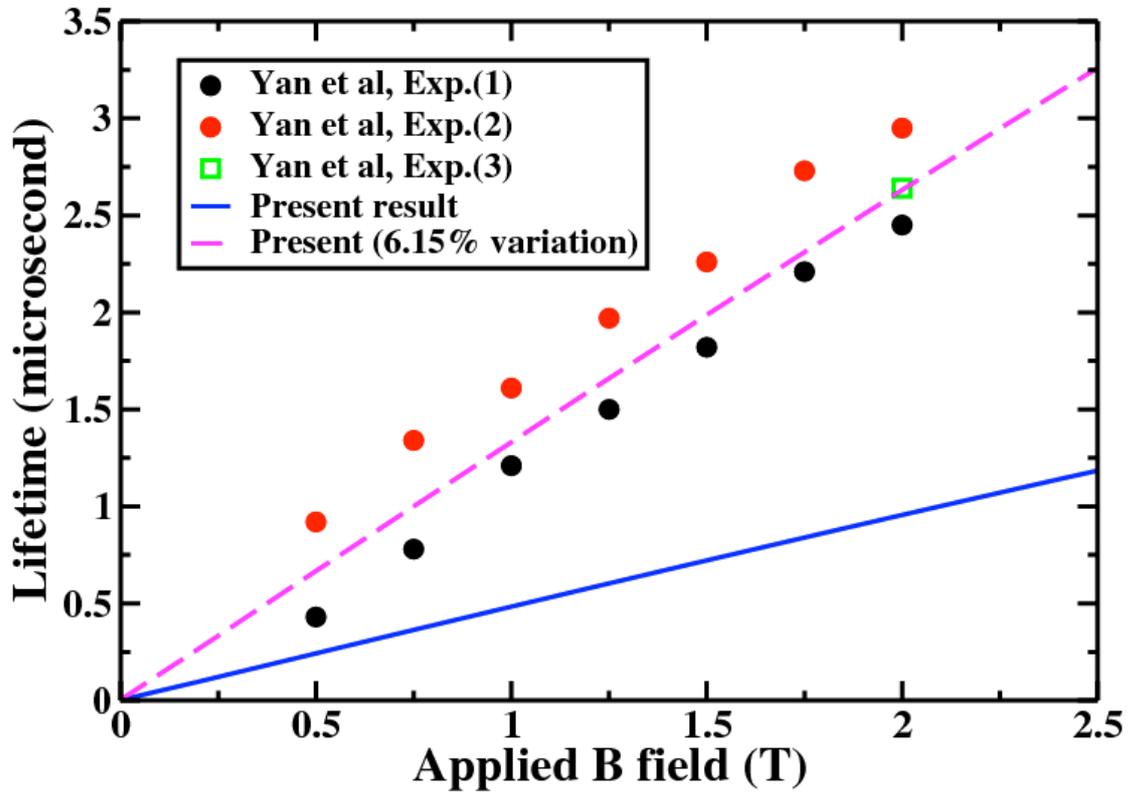

**Figure 4**: Comparison of the present results with experimental data by Yan et al[14] on Fe$_3$ chains. The full blue line shows the present results for the lifetime of the first excited state as a function of the applied B field. The dashed magenta line shows the present results when the magnetic structure parameters are modified by 6.15% (see text). Experimental data by Yan et al: measurements on central and edge Fe atoms (black and red dots), asymptotic 'unperturbed' value for a large tip-chain distance and B = 2T (green square).

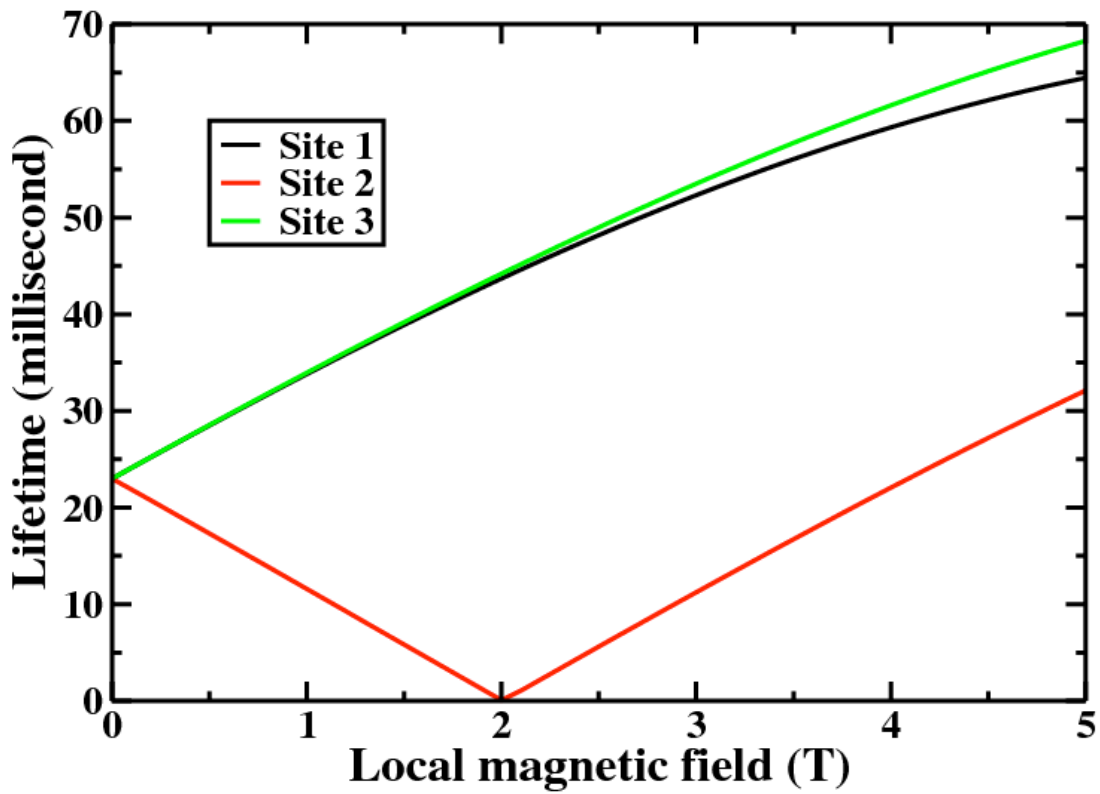

**Figure 5**: Lifetime of the first excited state (quasi Néel state $N_2$) of a $Fe_5$ chain, when a local magnetic field is applied on one of the atoms in the chain (site 1 is an edge atom, site 3 the central atom and site 2 an intermediate atom). The lifetime (in ms) is plotted as a function of the local field, $B_{Local}$, in T. A global magnetic field, $B$, equal to 2 T is also applied to the chain. As shown in Yan et al[14], introducing this local field mimics the perturbation induced by a polarized tip on the chain, so that the three curves represent the lifetime measured on the three atomic different sites in $Fe_5$ as a function of the perturbation introduced by the tip.